\documentclass[]{aastex631}
\usepackage{graphicx}
\usepackage{float}
\usepackage[verbose]{wrapfig}
\usepackage{url}
\usepackage{natbib}
\usepackage{color}
\usepackage{comment}

\def\beq{\begin{equation}}
\def\eeq{\end{equation}}

\def\Rs{$R_\odot$}

\def\f{{\it Fermi}\ }

\def\b{{\beta}}
\def\d{{\delta}}

\def\g{{\gamma}}

\def\l{{\lambda}}

\def\tsc{\tau_{\rm sc}\,}

\def\tcross{\tau_{\rm cross}\,}
\def\tesc{T_{\rm esc}\,}

\newcommand{\mj}[1]{{\color{black}#1}}
\newcommand{\vahe}[1]{{\color{black}#1}}

\shorttitle{Transport OF CME Protons to the Sun}
\shortauthors{Petrosian, Jin}

\begin{document}

\title{Transport of High Energy Protons from CME-Shock to the Sun and its Connection to Extended Gamma-rays Observed by {\it FERMI}}

\author{Vah\'e Petrosian}

\affiliation{Department of Physics  Stanford University, 382 Via Pueblo Mall, Stanford, CA 94305-4060}
\affiliation{Kavli Institute for Particle Astrophysics and Cosmology, Stanford University}
\affiliation{Department of Applied Physics, Stanford University}

\author{Meng Jin}
\affiliation{Lockheed Martin Solar and Astrophysics Lab (LMSAL), 3251 Hanover St., Bldg.203, Palo Alto, CA 94306, USA}

\email{vahep@stanford.edu}

\begin{abstract}

 This work is motivated by the \f Large Area Telescope (LAT)  observations of long-duration gamma rays from flares associated often with fast Coronal Mass Ejections (CMEs), several of which originate behind the limb (BTL) of the Sun. Our aim is to test the feasibility of the scenario whereby the gamma-rays are produced by CME-shock accelerated particles escaping from downstream to the Sun by investigating the transport of protons in this process for the flare of 2014 September 1. We use results from the global MHD modeling by the Alfv\'{e}n Wave Solar Model (AWSoM) to obtain the characteristics of downstream plasma, such as magnetic field geometry, Alfv\'en speed, and energy density of turbulence. This allows us to evaluate the variations of the proton scattering mean free path with  energy and distance from the CME-shock. Using an approximate analytic relation between the relevant time scales of the transport, we determine the energy dependence of the residence time at (or escape time from)  different distances above the photosphere, and  determine the evolution of  the proton spectrum from the acceleration site to the photosphere. These results can be used to obtain the spectrum of  the observed gamma-rays. Inversely, because of insignificant energy loss or gain during transport, this is a reversible process and the spectrum of protons at the photosphere, obtained from the observed gamma-rays, can be ``de-evolved" back to the CME using the same algorithm. Comparison of these spectra with the in situ observations of Solar Energetic Particles (SEPs)  could then shed light on their transport from the CME to the Earth.

\end{abstract}


\section{Introduction}
\label{sec:intro}

The Fermi Gamma Ray Observatory ({\it Fermi}) observes the Sun once every other orbit, and its Large Area Telescope (LAT) has detected $>100$ MeV photons from more than 50 solar flares, some with very long duration (extending up to a day). These flares are  often associated with faster than average Coronal Mass Ejections (CMEs) with speed $>$1000 km s$^{-1}$. Flares with sufficient flux show significant drift  of the emission  centroid  across the photosphere to areas far from the active region (AR) where the flare was initiated, and four originate from ARs a few to tens of degrees behind the limb (BTL) of the Sun \citep{ackermann2017}.  In this paper we explore the scenario where gamma-rays are produced by CME-driven shock accelerated ions (mainly protons) reaching the photosphere from the downstream of the shock. \mj{In this scenario, the protons are energized by the diffusive shock acceleration (DSA) mechanism in a thin region around the shock surface, where repeated shock crossings, mediated by scattering in the turbulent regions immediately upstream and downstream of the shock, produce the accelerated particle spectrum. The particles escaping the acceleration region, from upstream to $\sim 1$ au or downstream to the Sun  are scattered by turbulence but are not further accelerated. Scattering plays an important role in determining the flux and spectrum of the particles throughout their transport to $\sim 1$ au and to the photosphere.}

Global MHD modeling of CMEs shows that as CMEs propagate outward, magnetic fields connect them to regions far from the flare site \citep{jin18}. This connection helps explain the observed centroid drifts and provides some support for the scenario described above. \mj{An additional piece of observational evidence that has been put forward in support of this scenario is the reported correlation between the durations of gamma-ray emission and of type II radio emission, which is produced by electrons escaping from the upstream region of the CME-driven shock \citep{gopalswamy18}; we note, however, that the statistical significance of this correlation has been questioned (\citealt{bruno23}, see below).} These observations provide necessary but not sufficient evidence for the proposed scenario; establishing sufficiency requires a detailed investigation of the particle transport process, which depends on the geometry of the magnetic field ($B$ field) and the properties of the plasma downstream of the shock, particularly the characteristics of the MHD turbulence.

This scenario has faced some objections. For example, \citet{hudson18} and \citet{klein18} argued that the magnetic fields stretched and expanded by the CME would be highly convergent downward, resulting in a very small loss cone and, consequently, a limited fraction of protons reaching the photosphere. While this is true for freely moving particles, we will show that this is not the case when turbulence downstream of the shock is taken into account. Turbulence can scatter particles into the loss cone \mj{(e.g., \citealt{hutchinson2022})}, and, more importantly, when the scattering mean free path, $\l_{\rm mfp}$, is smaller than the magnetic field scale height, $H_B = |d\ln B/dr|^{-1}$, the field convergence has a limited effect. In the case where $\l_{\rm mfp}\ll H_B = |d\ln B/dr|^{-1}$, a large fraction of particles can reach the photosphere.

Several studies \citep{ryan00, denolfo19, bruno23} have also raised doubts about the validity of the correlation between type II radio emission and gamma rays. These works also point out that the temporal evolution of solar energetic particles (SEPs) escaping the CME toward Earth often differs from that of gamma rays, and that in some cases, the ratio of the number of protons required to produce the observed gamma rays to the number of detected SEPs can be high. \citet{denolfo19} obtain very high ratios, some exceeding 1. However, a more recent paper \citep{bruno23} shows ratios in the the range 0.4 to 0.1. These studies favor a model in which gamma-ray-producing protons are accelerated in large coronal loops. \mj{First, the reanalysis by \citet{bruno23} indicates that the duration correlation reported by \citet{gopalswamy18} is weaker than originally claimed and is largely driven by the few strongest events in the sample while \citet{gopalswamy21} have addressed some of the earlier criticisms. The statistical significance of the correlation across the full event sample thus remains an open question that warrants further investigation with larger samples and uniform analysis methods.} Second, direct comparisons between the number, and temporal evolution, of gamma-ray-producing and SEP protons observed at 1 au is not straightforward. The spectrum, flux, and temporal evolution of particles in the upstream and downstream regions in the immediate  surroundings of the shock are expected to be similar, because efficient acceleration by the shock requires  repeated both-way crossings of the shock. But the characteristics of particles escaping upstream to 1 au and those transported from the downstream to the Sun will be different significantly, owing to the different magnetic field geometries, turbulence levels, and magnetic connectivity in their respective transport environments. Furthermore, the observed centroid drifts and BTL flares argue against acceleration in coronal loops, as this would require evolving loops extending several solar radii and exhibiting a large foot point displacements, which are difficult to reconcile with current observations. \mj{Although such long loops have been observed in a recent event \cite{bruno2025}.} In addition, in the low beta plasma expected in the corona, a larger number of electrons than protons are expected to be accelerated \citep{petrosian04}. Signatures of electrons (mainly bremsstrahlung hard X-rays) are seen for the short impulsive phases only and  are originating from regions far from the centroids of gamma-ray emission \citep{ackermann2017}. However, recent \textit{Fermi} observations indicate that, for some long-duration gamma-ray events, a fast and strong CME-driven shock may not be a necessary condition \citep{pesce-rollins2024, bruno2025}.

In this paper, we extend the simulations by \citet{jin18} to investigate not only the magnetic field ($B$ field) structure but also key properties of the downstream plasma, such as the evolution of turbulence energy density and Alfv\'{e}n speed behind the shock as it propagates outward. These properties are essential for understanding the transport of particles to the Sun. In \S 2, we present results from AWSoM simulations detailing the evolution of these downstream characteristics as the CME-driven shock moves from $\sim$2 to $\sim$11 solar radii ($R_\odot$).%
\footnote{All the distances $r$ we use in this study are measured from the center of the Sun.}
In \S 3, we briefly review the particle transport equation and use the AWSoM results to evaluate the evolution of transport coefficients—specifically, the proton mean free path, $\l_{\rm mfp}$ (or the related scattering time, $\tsc=\l_{\rm mfp}/v$), and the associated residence or escape time, $\tesc$—as functions of particle energy. In \S 4, we apply these results to determine the evolution of the accelerated particle spectrum down to the photosphere. A brief summary and our conclusions are presented in \S 5.

\section{Plasma downstream of the CME-shock}
\label{sec:AWsome}

The AWSoM model is a global MHD model driven by observed magnetic maps, with its computational domain extending from the upper chromosphere to the heliosphere \citep{vanderholst14, sokolov13}, and is implemented within the Space Weather Modeling Framework (SWMF; \citealt{toth12}). AWSoM employs a phenomenological approach to Alfv\'{e}n wave dissipation, meaning the wave spectrum is not explicitly resolved. Instead, it calculates the total energy densities of counter-propagating Alfv\'{e}n waves. The energy dissipated by these waves heats the corona, while the resulting wave pressure gradient accelerates the solar wind. The model also treats electron and proton temperatures separately, with coupling via collisions, and uses adaptive mesh refinement to resolve fine structures such as current sheets and shocks. With its incorporation of key physical processes like Alfv\'{e}n wave dissipation and separate electron and proton temperatures, AWSoM is capable of reproducing high-fidelity solar corona conditions (e.g., \citealt{jin17a}, \citealt{sachdeva19}), thereby providing a realistic environment through which CMEs propagate.

\newcommand{\imsize}{0.8\columnwidth}
\begin{figure}
\begin{center}
{\resizebox{\imsize}{!}{\includegraphics{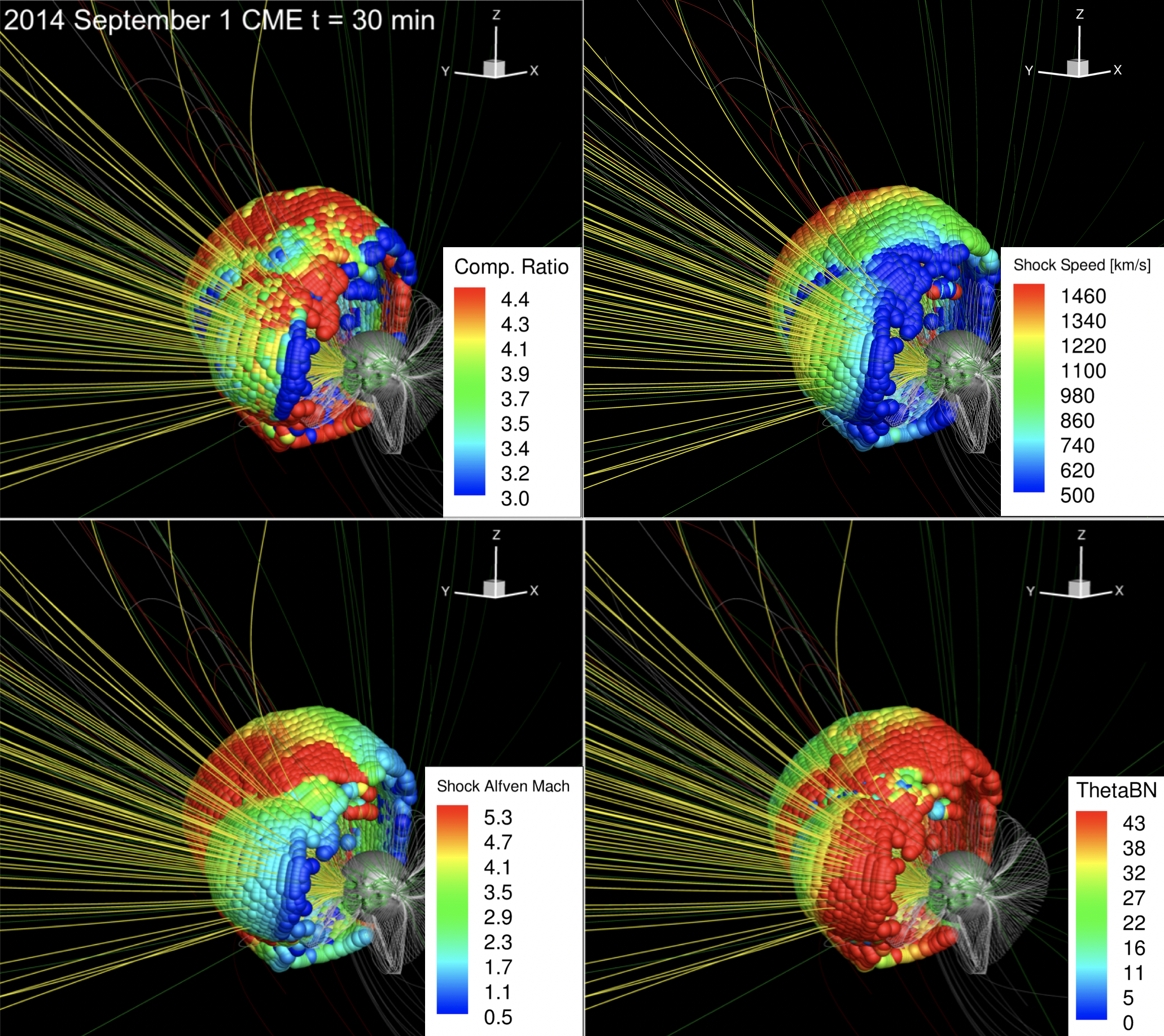}}}
\end{center}
\caption{CME-driven shock parameters (clockwise from top left; compression ratio, shock speed, The B field-shock velocity angle $\theta_{Bn}$), and shock Alfv\'{e}n Mach number)  at t = 30 min. The yellow field lines represent the open field near the \f-LAT gamma-ray emission region that connect to the shock. Adopted from \citet{jin18}.}
\label{mhd}
\end{figure}

To initiate a CME eruption, The Gibson–Low (GL) analytical flux rope model \citep{gl98} is used, which features a dense core at the bottom—representing filament material—and a density-depleted cavity surrounding the main flux rope body. A notable advantage of the GL flux rope model is its ability to reproduce the typical three-part density structure of CMEs observed in white light. To facilitate CME initiation, the Eruptive Event Generator Gibson–Low (EEGGL) module was developed to calculate flux rope parameters automatically based on the input magnetic map and desired CME speed, enabling applications in space weather forecasting \citep{jin17b, borovikov17}. In this study, we utilize the simulation data presented in \citet{jin18}; for detailed information on the model setup and event simulation, please refer to that paper.

\begin{figure}
		\centering
		\includegraphics[width=0.49\textwidth]{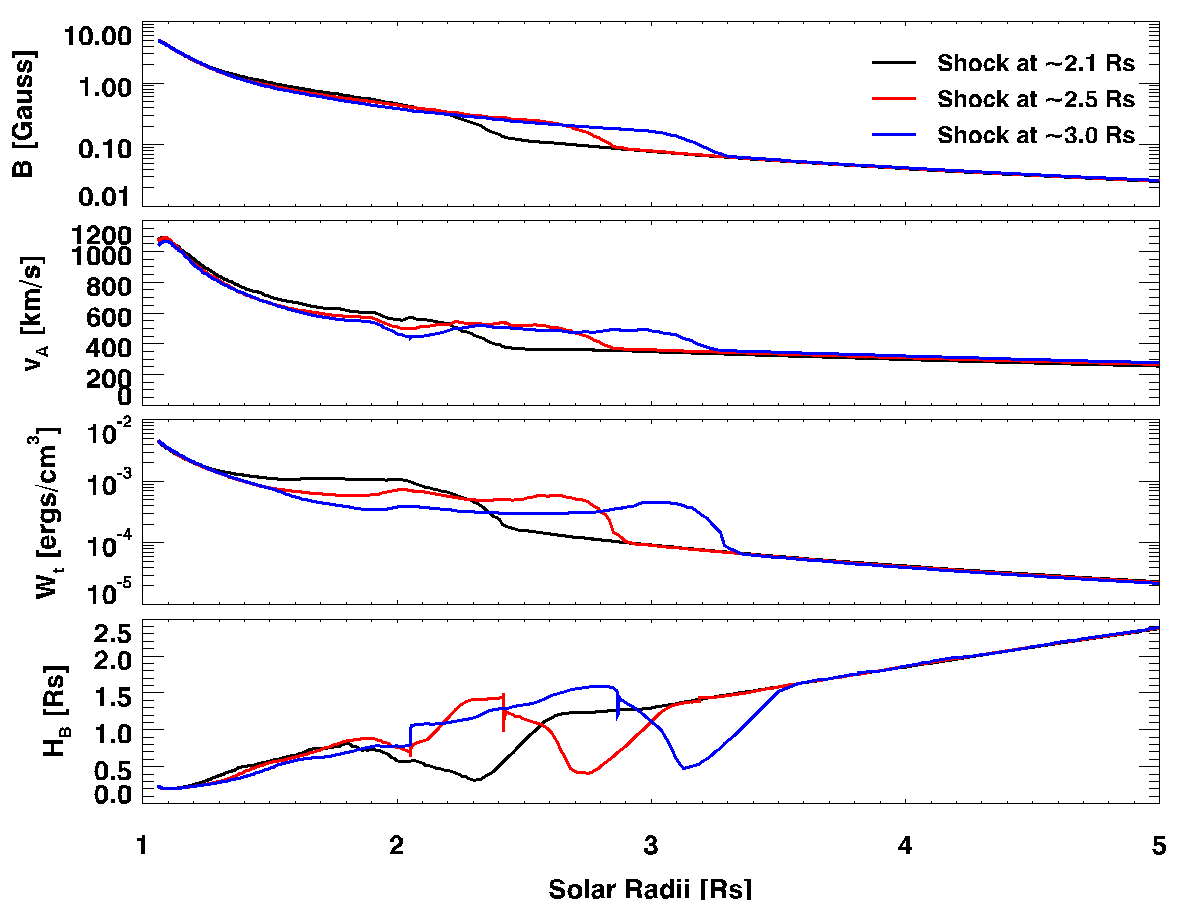}
		\includegraphics[width=0.49\textwidth]{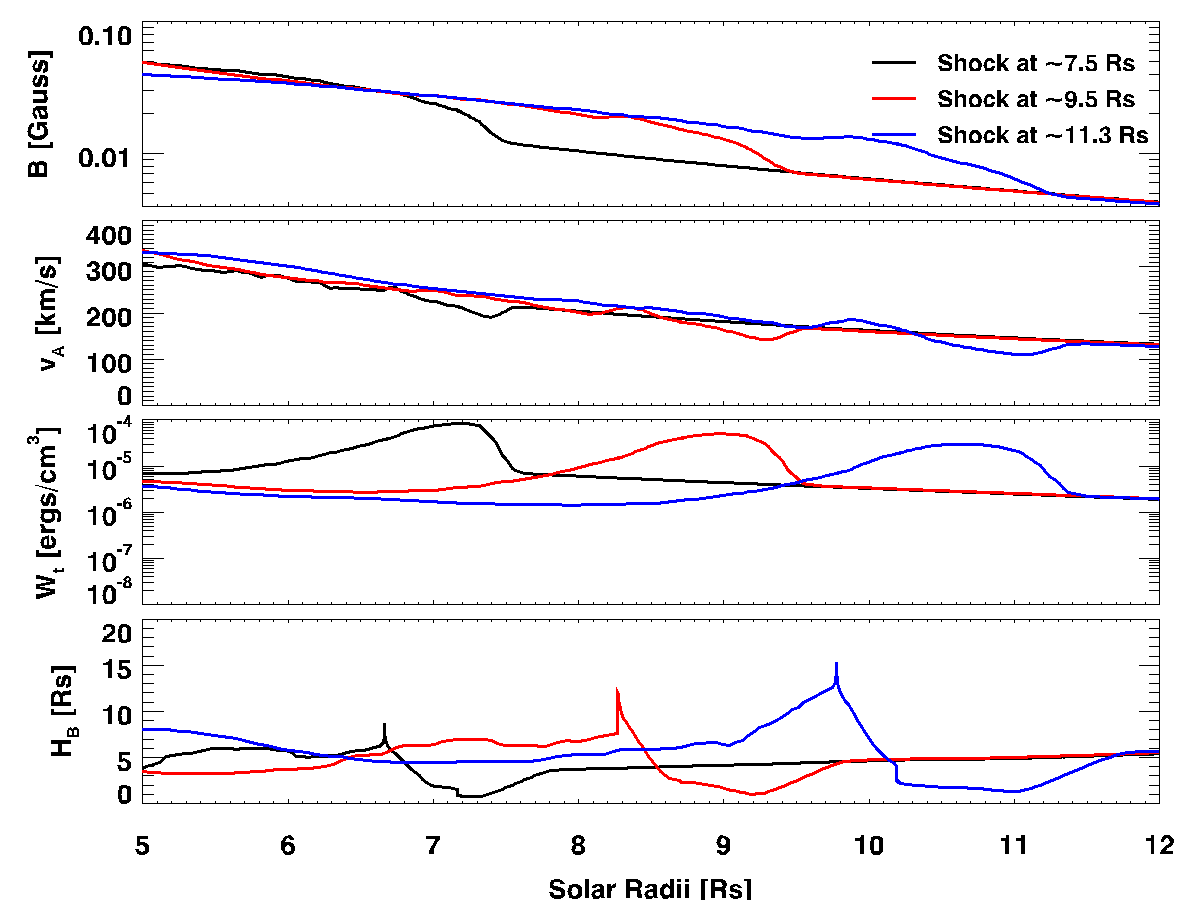}
		\caption{Variation of some key variables with distance in the downstream of the shock to the photosphere and upstream to $r \sim 12$ \Rs, for three shock distances (left: 2.1, 2.5 and 3.0 \Rs~from the center of the Sun; right: 7.5, 9.5, 11.3 \Rs~from the center of the Sun). From top to bottom;  magnetic field, $B$ (in Gauss), Alfv\'en velocity,  $v_A$ (in km/s), energy density of the Alfv\'enic turbulence, ${\cal {\cal W}_0}$ (in erg cm$^{-3}$), and $B$ field scale height, $H_B=|d\ln B/dr|^{-1}$ (in units of \Rs). Note that the outer boundary of the simulation is up to 24 \Rs.}
		\label{fig:shockparams}
	\end{figure}

From the simulation data, we extract variables from representative field lines connecting from the shock to the surface of the Sun within the Fermi gamma-ray emission centroid in the 2014 September 1 event (the yellow field lines marked in Figure~\ref{mhd}). Figure~\ref{fig:shockparams} shows the spatial and temporal profiles of several key variables—magnetic field strength, Alfv\'{e}n velocity, turbulence energy density, and magnetic scale height—relevant to the treatment of particle transport both upstream (in the solar wind) and downstream of the shock. These profiles are presented for three shock positions near the Sun: 2.1, 2.5 and 3.0 \Rs~from the center of the Sun (\textit{left panel} of Figure~\ref{fig:shockparams}), or at three different times after the CME onset; and three shock positions further away from the Sun: 7.5, 9.5, and 11.3 \Rs~from the center of the Sun (\textit{right panel} of Figure~\ref{fig:shockparams}). Our selection of distances for MHD simulation snapshots is guided by the Fermi observations of the 2014 September 1 event. In this event, the Fermi $>$100 MeV flux peaks approximately 20 minutes after the flare onset, which coincides with the peak in the simulated shock compression ratio (see Figure 8 in \citealt{jin18}). The CME-driven shock started to intersect the open field lines around t = 6 minutes in the simulation, when the shock is around $\sim$1.6 R$_\odot$. At this peak time, the CME-driven shock intersecting the open magnetic-field lines that connect to the visible solar disk is located at roughly 3 R$_\odot$. Fermi-LAT detects the flare at the second passing of the Sun (90 minutes later) but with a two orders of magnitude weaker flux. We extend the MHD simulation to pass 100 minutes when the CME-shock is at $>$11 \Rs~to have the full coverage of the Fermi-LAT observation.

\begin{figure}
		\centering
		\includegraphics[width=0.46\textwidth]{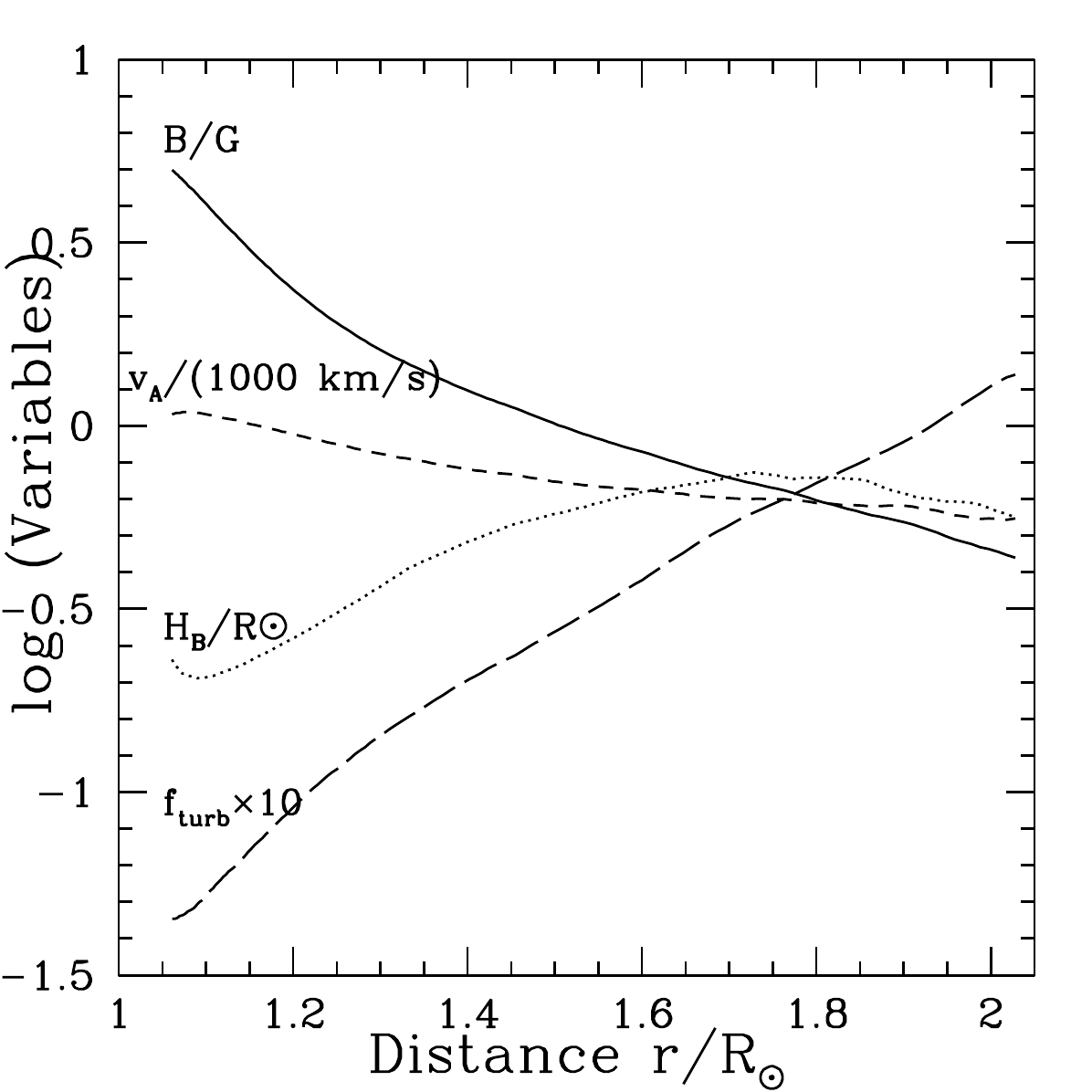}
		\includegraphics[width=0.478\textwidth]{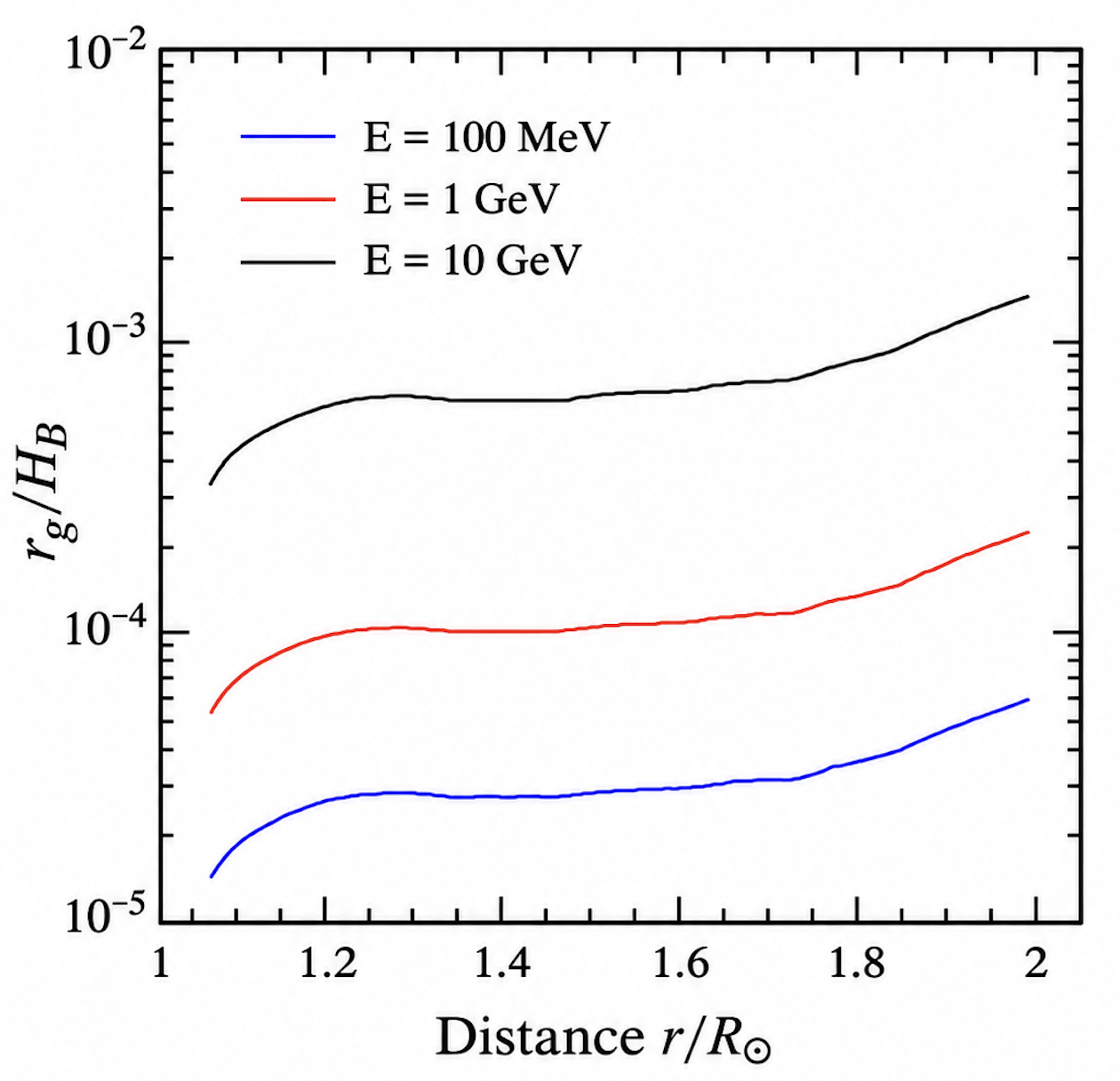}
		\caption{{\it Left:} variation with distance from the center of the Sun to the CME-shock of the magnetic field, $B$/G,  its scale height, $H_B=|d\ln B/dr|^{-1}$, Alfv\'en velocity in units of  $10^3$ km/s,  and the fraction of energy in Alfv\'enic turbulence, $f_{\rm turb}=8\pi{\cal W_0}/B^2$ multiplied by 10. {\it Right:} Variation with distance (from the photosphere  to the CME-shock at $r \sim 2$ \Rs) of the ratio of the proton gyroradius to magnetic field scale height, for three different energies.}
		\label{fig:params}
	\end{figure}

Using the above results from the AWSoM simulation, in Figure~\ref{fig:params} (left) we show the spatial variation of the above variables in the downstream of a shock at distance $r\sim 2$\Rs. As mentioned before, the AWSoM MHD model uses phenomenological approach to Alfv\'{e}n wave dissipation, in which there are two wave kinetic equations included in addition to the MHD equations and the counter-propagating wave turbulence energy densities are self consistently calculated in the model. The total turbulence energy density can be derived as ${\cal W}_0$ = ${\cal W}_-$ + ${\cal W}_+$, where ${\cal W}_-$ and ${\cal W}_+$ are the direct outputs from the AWSoM model.
Instead of turbulence energy density, ${\cal W}_0$, we plot the fraction of magnetic energy in turbulence,  $f_{\rm turb}=8\pi {\cal W}_0/B^2$.

\section{Review of Transport and its Coefficients}
\label{transport}

The exact treatment of the transport of energetic particles in strongly magnetized, turbulent plasma requires numerical solution of the Fokker-Planck kinetic equation describing the particle phase space distribution $f(t,\vec{r}, {\vec p})$, as a function of time, distance, and momentum ${\vec p}$, and involves advection, effects of the large scale magnetic field variation, several diffusion coefficients arising from scattering by turbulence, and energy loss. However, for our problem, namely the transport of protons from  the CME-shock downstream region to the Sun, this complex equation is simplified considerably.

\begin{enumerate}
    \item

As shown below, the dynamic scale time determined by the velocity of the CME  or slower solar wind (from minutes to hours) is longer than most of the relevant transport times. Thus, we can assume a steady state situation and ignore the time dependence.

\item

In a strong guiding field, the particle gyro-radius, $r_g$,  is shorter than the source size, or scale height $H_B$. Thus, the phase space distribution simplifies to $f(s,\mu,E)$, where $s$ is the distance along the field line, $\mu$ is the cosine of the pitch angle and $E$ is the kinetic energy. Gyro-radius of protons (with velocity $v$ and Lorentz factor $\g=1+E/(mc^2)$) is $r_g = v\g\sqrt{1-\mu^2}/\Omega_p$, where $\Omega_p = eB/(m_pc) =
9.6\times 10^3 (B/$G) is the proton gyro-frequency. For an isotropic distribution of pitch angles the average value of $\sqrt{1-\mu^2} = \pi/4$, and we obtain
$r_g = 3.5\times 10^{-5}\sqrt{\g^2-1}({\rm G}/B)$\Rs, which, as shown in Figure \ref{fig:params} (right), is much smaller than $H_B$.

\item

\mj{After leaving the acceleration region at the shock protons gain negligible energy because the timescale of  stochastic acceleration by the downstream turbulence is much larger than the scattering and residence (or escape) times calculated below. They also lose insignificant energy until they penetrate the photosphere where they produce gamma-rays.}
Under these conditions, the Fokker-Planck equation simplifies to
\beq
 v\mu{\partial f\over \partial s}
+ {v \partial {\rm ln} B\over 2 \partial s}
{\partial\over\partial\mu}\left[(1-\mu^2)f\right]
- {\partial \over \partial \mu} \left( D_{\mu\mu} {\partial f\over \partial
\mu}\right)
={\dot Q},
\label{FPeq}
\eeq
where $D_{\mu\mu}$ is the pitch angle diffusion rate due to proton-turbulence scattering, and $\dot {Q}(\mu, E)$ is the injected source spectrum  at $s=0$.%
\footnote{A similar equation, with possibly different ${\dot Q}$, describes the transport of the energetic particles escaping the upstream of the shock to 1 au.}

This equation involves the following spatial (or time) scales: distance $\Delta r$ of the CME from the photosphere (or crossing time $\tcross=\Delta r/v$), magnetic field  scale height $H_B=|d\ln B/dr|^{-1}$ (or $\tau_B=H_B/v$), and scattering mean free-path $\l_{mfp}$, or scattering time

\beq
\tsc=\l_{mfp}/v=(1/2)\int_{-1}^{+1} d\mu (1-\mu^2)/D_{\mu\mu},
\label{tscat1}
\eeq
the detail of which is described next.

\end{enumerate}

\subsection{Scattering Time}
\label{sec:tscat}

The pitch angle diffusion coefficient, and hence the scattering time, is determined by the wave-particle interactions, which depend on several variables and parameters; magnetic field, $B(r)$, plasma density, $n(r)$, or their combination the  Alfv\'en velocity, $v_A$, a measure of the degree of magnetization, and spectrum of turbulence energy density, ${\cal W} (k,r)$, as a function of the wavevector $k$. In the quasilinear method  the gyro-resonance  interaction times (or rates), including $D_{\mu,\mu}$, scale with the characteristic time, $\tau_p$, or its inverse the characteristic interaction rate, given by the following expression for protons [see \citealt{dung94, pryadko97, pryadko99, petrosian04, jiang09}]

\beq
\label{taup}
\tau_{p,p}^{-1}=(\pi/2)\Omega_p f_{\rm turb}(q-1)(ck_{\rm
min}/\Omega_p)^{(q-1)},
\eeq
where $f_{\rm turb}=(8\pi {\cal W}_0/B^2)=(\d B/B)^2$ is the ratio of the total energy density of turbulence to that of the total $B$ field. This rate and the  Alfv\'en velocity are the two most important variables. AWSoM provides the Alf\'ven velocity, the magnetic field $B$ and  the total energy density of the turbulence, ${\cal W}_0$, but not its spectrum, ${\cal W} (k)$. However, it is reasonable to assume that in the inertial range, $k_{\rm min}<k<k_{\rm max}$, the spectrum follows a power law, ${\cal W}(k,r)dk={\cal W}_0(r)\times (q-1)(k/k_{\rm min})^{-q}(dk/k_{\rm min})$, with $q=5/3$ or 3/2  for the two most likely models of turbulence; Kolmogorov and Iroshnikov–Kraichnan (IK), respectively.%
\footnote{Parker Solar Probe indicates that the solar wind turbulence obeys Kolmogorov form from 1 au to $r\sim 0.2$ au and IK for $0.1<r<0.2$ au \citep{raouafi23}.}
\vahe{Also unknown is $k_{\rm min}$, the wave number of injected turbulence, a measure of the largest scale of the ``correlation" of turbulence. This scale is expected to be shorter than the ``size of the source" $\sim r$ or $H_B$. In what follows, we set $k_{\rm min}=1/(\xi H_B)$, with $\xi< 1$ as a free parameter. An estimate of $\xi$ can be obtained from Parker Solar Probe (PSP) observations, which show that the turbulence spectrum flattens to $q\sim 1$ below a frequency $f_{\rm min}$ that varies slowly with heliocentric distance $r$. If this break marks the outer scale of the inertial range, then
$k_{\rm min}=2\pi f_{\rm min}/v_A$. Between $r$=0.7 to 0.3 au, $f_{\rm min}$ increases from 2 to 5 mHz, roughly as $f_{\rm min}\sim 1\times  r^{-1.3}$ mHz, while $v_A\sim 500(r/20 R\odot)^{-1.2}$ \citep{petrosian23}, so that $k_{\rm min}$ is nearly independent of $r$. In this range, $k_{\rm min}H_B\sim 3\times 10^3$, i.e., $\xi \sim 3\times 10^{-4}$. There are, however, no direct  measurements in the inner heliosphere region ($\leq 20R_\odot$), so $\xi$ remains uncertain, and we treat it as a free parameter and adopt $\xi=0.01$ in what follows.}

\vahe{A further uncertainty concerns the energy densities of the turbulence and of the magnetic field, and hence their ratio $f_{\rm turb}$, in the corona and inner heliosphere ($r \lesssim 0.1$ au). PSP has, however, provided data on both quantities at larger distance ($0.1 <r <1$ au). As shown in \citet{petrosian23}, these and near-Earth observations are well described by the power laws $B(r)=B_0(r/R_\odot)^{-\d}$, with $B_0\simeq 1.0$~G and $\d\simeq 1.9$, and ${\cal W}(r)={\cal W}_0(r/R_\odot)^{-3.1}$ with ${\cal W}_0=0.033~{\rm G}^2$, which, extrapolated inward, yield:}
\beq
f_{\rm turb}(r)=0.033(r/R_\odot)^{0.7}, \,\,\,{\rm for} \,\,\, 0.1~<r <1 {\rm au}.
\label{fturb}
\eeq

\begin{figure}
		\centering
		\includegraphics[width=0.45\textwidth]{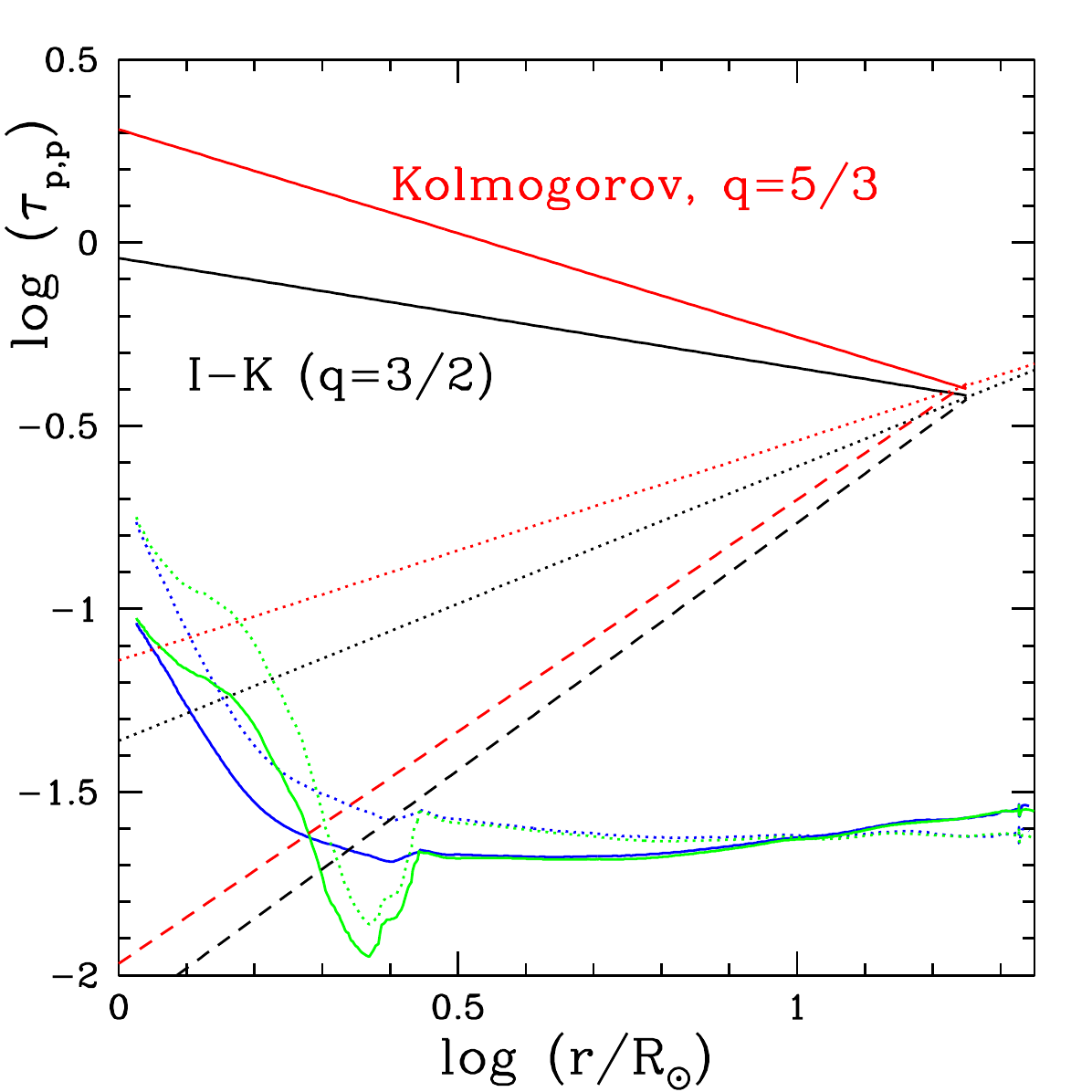}
		\includegraphics[width=0.46\textwidth]{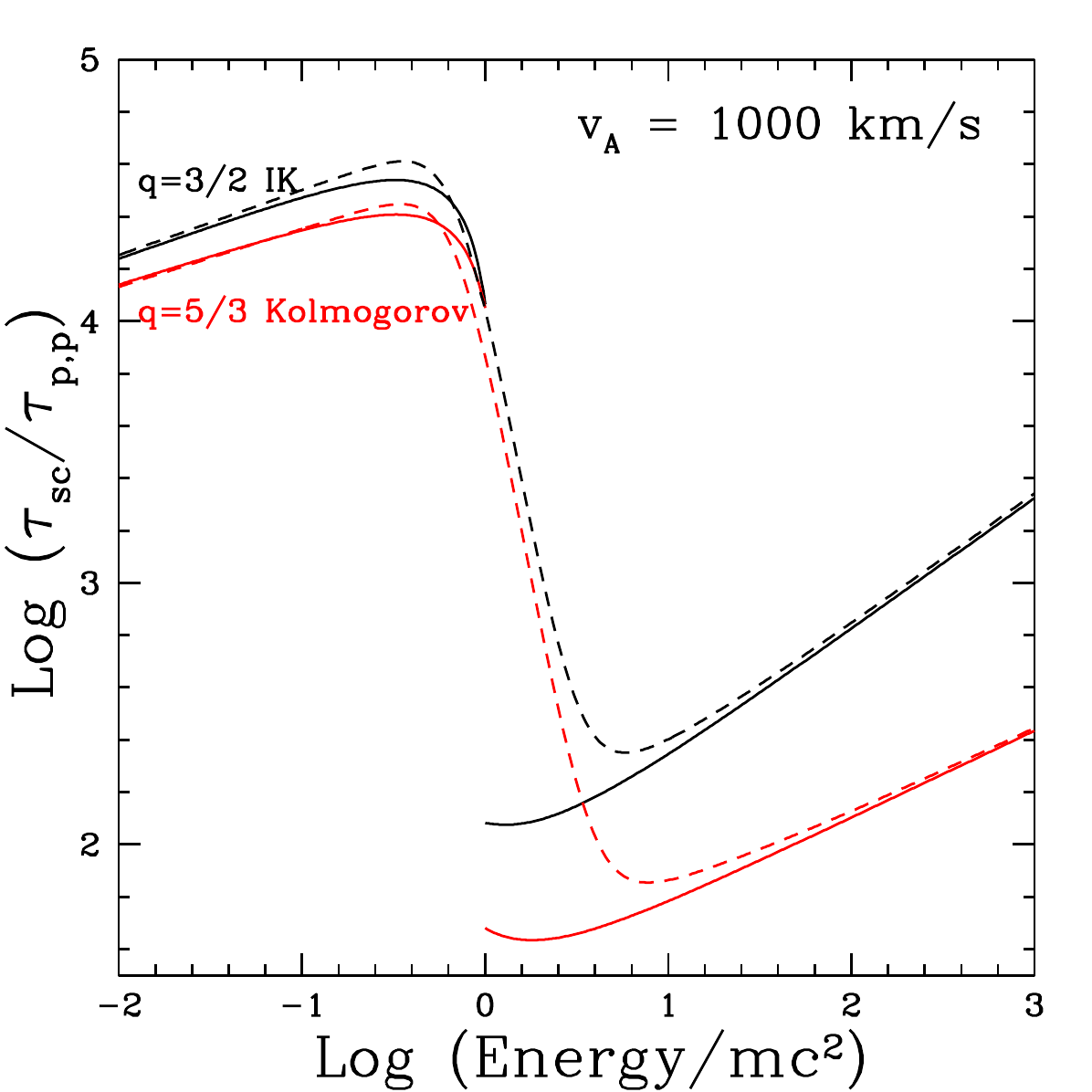}
		\caption{{\it Left:} Variation of $\tau_{p,p}$ with distance based on PSP observations at the outer heliosphere (dotted lines) with three interpolations (solid-low $B$; dashed-high $B$; dotted-direct interpolations) to the Sun for Kolmogorov and IK turbulence models. \mj{The result based on AWSoM simulations is shown by blue (background) and green (with a CME-shock at 2\Rs) curves with Kolmogorov (dotted) and IK (solid) turbulence model.} {\it Right:} Scattering time of protons as a function of energy (in units of $\tau_p$, and for two values of the turbulence spectral index $q$), in the relativistic regime (right), based on Equation (\ref{tscatRel}) extended to semi-relativistic regime using  $(\g\b)^n=E^n(1-2n/E)$, and in the  non-relativistic regime, based on Equation (\ref{tscatNonRel}), extended to semi-relativistic regime using $\b=\sqrt{2E}(1-3E/4)$; $\b_A=3.33\times 10^{-3}$. The dashed lines show the analytic fit based on Equation (\ref{tscfits}).}
        \label{fig:tauppandscattimes}
        	\end{figure}

\vahe{However, there are limited measurements of these quantities in the inner region. For example,  as shown in \citet{petrosian23} there is about one order of magnitude uncertainty in the value of the magnetic field at the Sun. In what follows  we extend the above power law for turbulence into the inner region and obtain a range of estimates of $\tau_{p,p}$. Rewriting $\tau_{p,p}^{-1}$ in terms of fitted quantities, we have:}
\beq
\label{taup2}
\tau_{p,p}^{-1}=(\pi/2)(q-1)[\d c/(\xi R_\odot)]^{(q-1)}\Omega_{p,0}^{(2-q)}({\cal W}_0/B_0^2)(r/R_\odot)^{(\d-1)q-2.1},
\eeq
\vahe{where $c$ is the speed of light and $\Omega_{p,0}=eB_0/(m_pc)$.
The resultant variation of $\tau_{p,p}$ is shown by the power-law forms in the left panel of Fig.~\ref{fig:tauppandscattimes}, for $\d=2.6, B_0=8.4; \d=1.9, B_0=1.0$, and $\d=1.5, B_0=0.3$ from top to bottom, taken from \citet{petrosian23}. In what follows we will use the simple interpolation of the outer observations, the dotted lines ($\d=1.9, B_0=1.0$); $\tau_{p,p}\propto (r/R_\odot)^{(2.1-0.9q)}$;
along with the green and blue curves that show the expected variations based on the simulation result of AWSoM for the pre-CME background and with the CME-shock at 2 \Rs, respectively.}

In general, $D_{\mu\mu}$ and hence the scattering time, in addition to this scaling, depend, in a complicated way, on  $q$ and  $\beta_A=v_A/c$, and on particle energy and pitch angle (see, e.g. \citealt{pryadko97, pryadko99, petrosian04, jiang09}). However, for relativistic protons, $E>$ GeV, the primary interactions occur with Alfv\'{e}n waves, with the dispersion relation $\omega=k_\| v_A$ for waves propagating parallel and anti-parallel to the $B$ field. Interactions also occur with fast mode waves, where for  $k\ll 1$ the dispersion relation $\omega=kv_A$, is applicable to both parallel and perpendicular wave propagation. In this case, the equations describing the above characteristics are simplified.

As shown in \citet{pryadko97} and \citet{petrosian04} for parallel propagating waves with $q<2$, and   for relativistic protons with isotropic pitch angle
distributions the scattering time is given by%
\footnote{There is an error in Equation (C3) of \citet{petrosian04}. It contains the extra term $(w_p=m_e/m_p)^q$.}
\beq
\label{tscatRel}
\tau_{sc}=\tau_{p,p}\g^{(2-q)}\b^{1-q}{2\over (2-q)(4-q)}A(\b_A/\b,q),
\eeq
and
\beq
\label{Aq}
A(\b_A/\b,q)=1-(2-q/2)(\b_A/\b)^{2-q}+(1-q/2)(\b_A/\b)^{4-q}.
\eeq
For $\b_A\ll\b=v/c$, which is the case here, $A(q,\b_A)\simeq 1$.

The lower energy protons interact with other plasma waves complicating the results \citep{pryadko97, pryadko99, petrosian04}. As shown in \citet{petrosian04} for non-relativistic protons, the scattering time depends on
$\beta_A$ and proton velocity $v=c\beta$;
\beq
\label{tscatNonRel}
\tau_{sc}=\tau_{p,p}\b^{(1-q/3)}\b_A^{-2q/3}[q(1+q/6)]B(\b_A/\b, q).
\eeq
with
\beq
\label{Bq}
B(\b_A/\b,q)=[1+a(q)(\b_A/\b)^{1/3}+b(q)(\b_A/\b)^{2/3}]^{-1},
\eeq
where $a(q),\, b(q)<0.05$ are slowly varying functions of $q$, so that for $\b_A\ll\b$ we can set $B(\b_A,q)=1$.

The solid curves in the right panel of Figure \ref{fig:tauppandscattimes} show the variation of the ratio $\tsc/\tau_{p,p}$ for the above two limiting cases (non-relativistic and relativistic branches), extended approximately to the semi-relativistic regime. These curves are fit with the following approximate analytic equation that consists of two broken power law forms:
\beq
\label{tscfits}
\tau_{sc}=\tau_{p,p}\left[q(1+q/6)\b_A^{-2q/3}{(2E)^{(3-q)/6} \over 1+(E/0.7)^4} + {2\over (2-q)(4-q)}E^{(2-q)}\sqrt{1+(3/E)^{2(2-q)}}\right],
\eeq
with two fitting parameters, energy break values,  $E_1=0.7$ and $E_2=3$ (in units of $mc^2$). The dashed lines in Figure \ref{fig:tauppandscattimes}  (right) show the fit $\tsc/\tau_{p,p}$ for protons in the energy range $0.01<E<10^3$, and two values of the turbulence index $q$.

\begin{figure}
		\centering
		\includegraphics[width=0.98\textwidth]{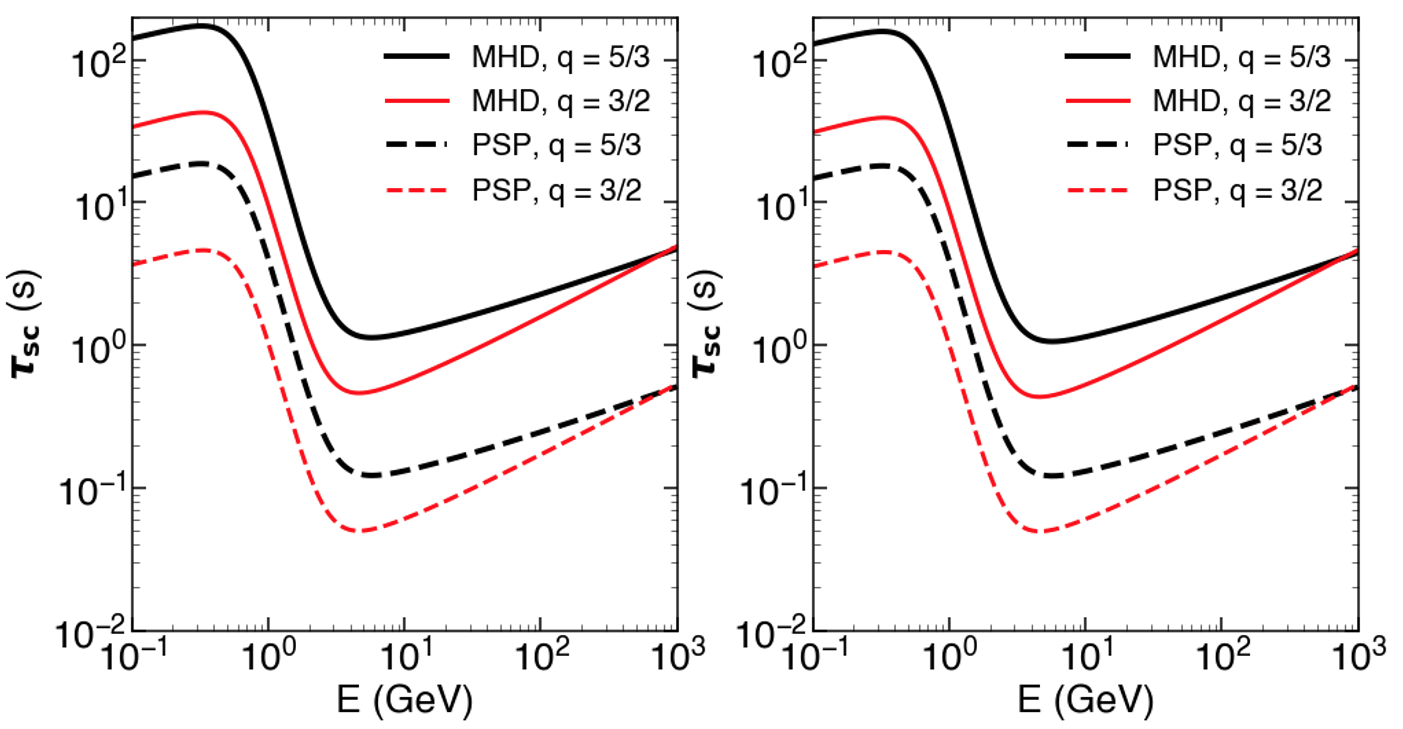}
		\caption{\mj{Scattering time of protons from Eq.~(\ref{tscfits}) for CME-shock at 2 \Rs (left) and at 11 \Rs (right). Solid curves use the characteristic time $\tau_{p,p}$ obtained from the AWSoM MHD simulation; dashed curves use $\tau_{p,p}$ from the power-law extrapolation of the PSP data (the middle, dotted curve in the left panel of Fig.~\ref{fig:tauppandscattimes}). Black curves: Kolmogorov turbulence ($q=5/3$); red curves: IK turbulence ($q=3/2$).}}
		\label{fig:scat1}
	\end{figure}

\mj{Substituting $\tau_{p,p}$ into Eq.~(\ref{tscfits}) gives the dependence of $\tsc$ on both energy and distance for the two turbulence models: the distance dependence enters entirely through the characteristic time $\tau_{p,p}$, and the energy dependence through the bracketed function in Eq.~(\ref{tscfits}). Since we are mainly interested in the transport near the Sun, Figure~\ref{fig:scat1} shows $\tsc(E)$ at the solar surface for both estimates of $\tau_{p,p}$, at the times when the shock is at 2~\Rs\ (left) and 11~\Rs\ (right). The PSP-based curves are identical in the two panels by construction, because the extrapolated power laws depend on distance only; the MHD-based curves are also nearly identical, because the plasma parameters near the Sun change little as the CME and its shock move outward. The shock distance therefore affects the transport not through $\tsc$, but through the crossing and escape times discussed next.} Note that at E = 400 MeV $\tsc$ peaks at values ranging from 150 (MHD/Kolmogorov model) to 4 s (PSP/IK model), yielding  the maximum mean free path $\lambda_{mfp}$ ranging from $\sim 50$ to 1 \Rs.

\subsection{The Escape Time}
\label{sec:escape}

\begin{figure}
		\centering
        \includegraphics[width=0.98\textwidth]{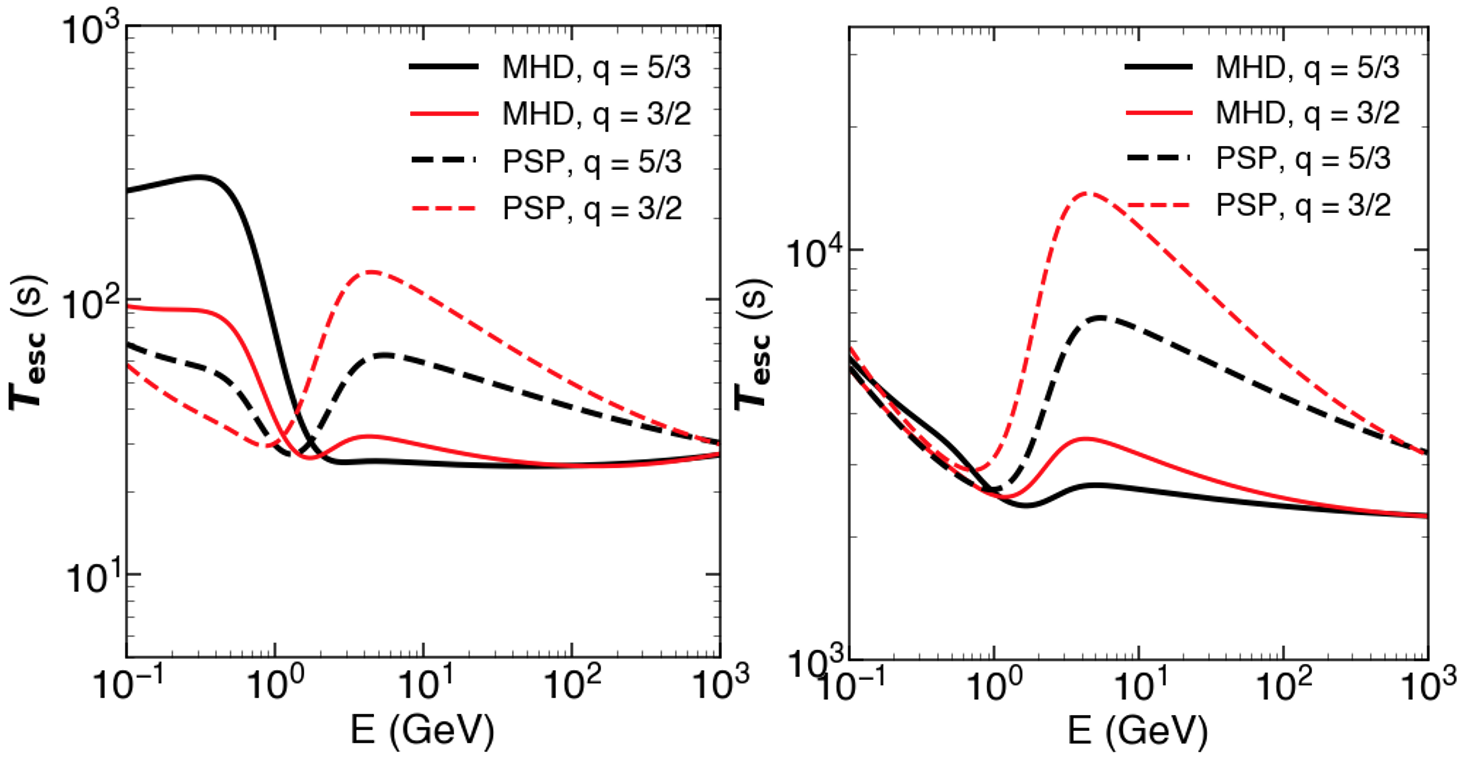}
		\caption{The escape time derived from Eq.~(\ref{tescape}) for same model parameters as in Fig.~\ref{fig:scat1} and \mj{for two shock distances at $r\sim2.1$ (left) and $r\sim11.3$ \Rs (right).}}
		\label{fig:tesc}
	\end{figure}

For an isotropic pitch angle distribution, we can  use pitch-angle integrated  quantities; namely the total density $N(s, E)=\int_{-1}^1 d\mu\, f(s,\mu,E)$ and the total injected rate from behind the shock ${\dot Q}(E,r_{\rm sh})=\int_0^1d\mu \, {\dot Q}(\mu, E,r_{\rm sh})$, leading to the well-known leaky box equation, where the concept of the escape time, $\tesc$, is commonly used (see,~e.g. review by \citealt{petrosian12}). In the absence of acceleration or energy loss, it is easy to show that this yields the following simple result:%
\footnote{This is similar to transport of galactic cosmic rays in the inner heliosphere described recently by \citet{petrosian23}.}
\beq
\label{spec}
N(r, E)=\tesc(r, E)\times {\dot Q}(E, r_{\rm sh}),
\eeq
where $\tesc(r,E)$ \mj{is the length of the time the particles with energy $E$ spend at any heliocentric distance $r$ in the downstream of the CME-driven shock. We refer to it as the residence time; the longer this time the higher the density of particles. This time also determines both the delay and duration of the gamma-ray emission relative to the injection at the shock, and the modification of the accelerated spectrum during the transport. It is related to the other time scales ($\tsc, \tcross$) and the magnetic field geometry ($H_B$).} The exact formulation of this process is complicated. However, in the strong diffusion limit, i.e.~when $\l_{\rm
mfp}
\ll  H_B=r/\d$, or equivalently when $\tsc=\l_{\rm mfp}/v \ll \tau_B\sim \tcross/\d)$, the pitch angle distribution will remain
isotropic, and we are dealing with the well-known random walk process, with
$\tesc\sim \tcross^2/\tsc\gg \tcross$, independent of  $B$ field convergence factor, $\eta\equiv r/H_B$.
In the opposite weak diffusion limit ($\tsc\ll \tcross\sim \tau_B$), the
particles are reflected and can escape toward the Sun only when scattered into the loss cone. Thus, the escape time $\tesc\propto \tsc$.
The numerical solution of Equation (\ref{FPeq}) by \citet{effenberger18}, give the relation between $\tesc$ and $\tsc, \tcross$ and $H_B$, which
fits well with the following relation suggested by \citet{malyshkin01} for the isotropic pitch angle scattering rate:%
\footnote{\citet{effenberger18} and \citet{malyshkin01} in their treatment of transport equation include only the effects of magnetic field mirroring and the pitch angle diffusion coefficient $D_{\mu\mu}\sim 1/\tsc$ for conditions very similar to transport in the downstream of CME-shock.}
\beq
\label{tescape}
\tesc/\tcross = \tcross/\tsc + 2\eta + \ln{\eta}(\tsc/\tcross).
\eeq
The first and third terms describe the above two limiting cases and are connected by a nearly constant (independent of the scattering time) middle term. \vahe{Note that $\tcross\sim (2/\b)(r/$\Rs)~s, where $\b=v/c>0.5$ for $E>0.1$~GeV, so that $\tcross$ varies from 4 to 2 s near the Sun, indicating that with increasing energy the transport passes from the weak to strong diffusion regime, especially for PSP models. Thus, we can calculate the variation of the escape or residence time, $\tesc(E)$, with distance and energy, shown in Figure~\ref{fig:tesc} for shock at $r\simeq 2.1$ (left) and $r\simeq 11.3$ \Rs (right). In contrast to $\tsc$, there is significant dependence on CME-shock distance from the Sun.}

\section{Proton Spectral Variation}
\label{sec:spectrum}

From Equation (\ref{spec}), we can calculate the expected proton spectral form, and its variation with distance, for a given injected spectrum, ${\dot Q}(E)$. The main information on this comes from the spectrum of SEP protons observed by instruments located at $\sim 1$~au. There is considerable spectral and composition variation in this data. However, for long duration events associated with CMEs, the spectra obey a broken power law form, sometimes modeled by the smoothly broken power-law Band function, with a power-law index $\d_p\sim 3$ to 4 at above 1 MeV, that flattens to $\d_p\sim 1$ for $E< \sim 1$ MeV. Production of \f-LAT gamma rays extending from 10 MeV to few GeV, requires protons in the energy range of 50 MeV to $\geq 5$ GeV. The directly observed SEP spectra extend to $\sim 100$MeV. There is indirect evidence for presence of $E\leq 1$~GeV protons in flares with ground level enhancements (GLEs), but no information on spectra. In addition, the exact relation between the SEP spectra observed at 1~au and those accelerated at the CME-shock environment at distances of less than 10 \Rs~is complicated and beyond the scope of this paper. However, something like the following scenario may be at work.

Around the shock, particles undergo repeated passages up and down through it, requiring short mean free paths and strong diffusion at the upstream and downstream regions near the shock. Some of the {\it upstream particles}  go back to the downstream, and some escape into the solar wind and eventually to 1 au. Similarly, some {\it downstream particles} cross the shock to the upstream, and some escape down toward the Sun. When the dynamic time, $t_{\rm dyn}\sim r/v_{\rm CME}$, is longer than the back-and-forth diffusion time, we expect similar density of accelerated particles just above and below the shock. However, because the  characteristics of turbulence in the upstream and downstream are most likely different, we expect different escape times, and thus a different spectrum of escaping particles from the upstream into the solar wind and from downstream to the Sun. In addition, these spectra are also altered differently during their transport to 1~au and the Sun, because of the difference in the characteristics of magnetic field and turbulence upstream  and downstream of the shock  region.%
\footnote{\mj{We do not model the DSA process itself. Its outcome is represented by the injected spectrum ${\dot Q}(E)$ released just behind the shock.}}
In what follows we assume a power law with exponential cutoff at high energies for protons escaping the downstream region of the CME-shock:
\beq
{\dot Q}(E,r_{sh})= {\dot Q_0}(r_{sh})E^{-\d_p}\times e^{-E/E_c},
\,\,\,\, {\rm with} \,\,\,\, \d_p\sim 3-5, \,\,\,\, {\rm and} \,\,\,\, E_c\leq 1~{\rm GeV}.
\label{injectedspec}
\eeq

\begin{figure}
		\centering
        \includegraphics[width=0.98\textwidth]{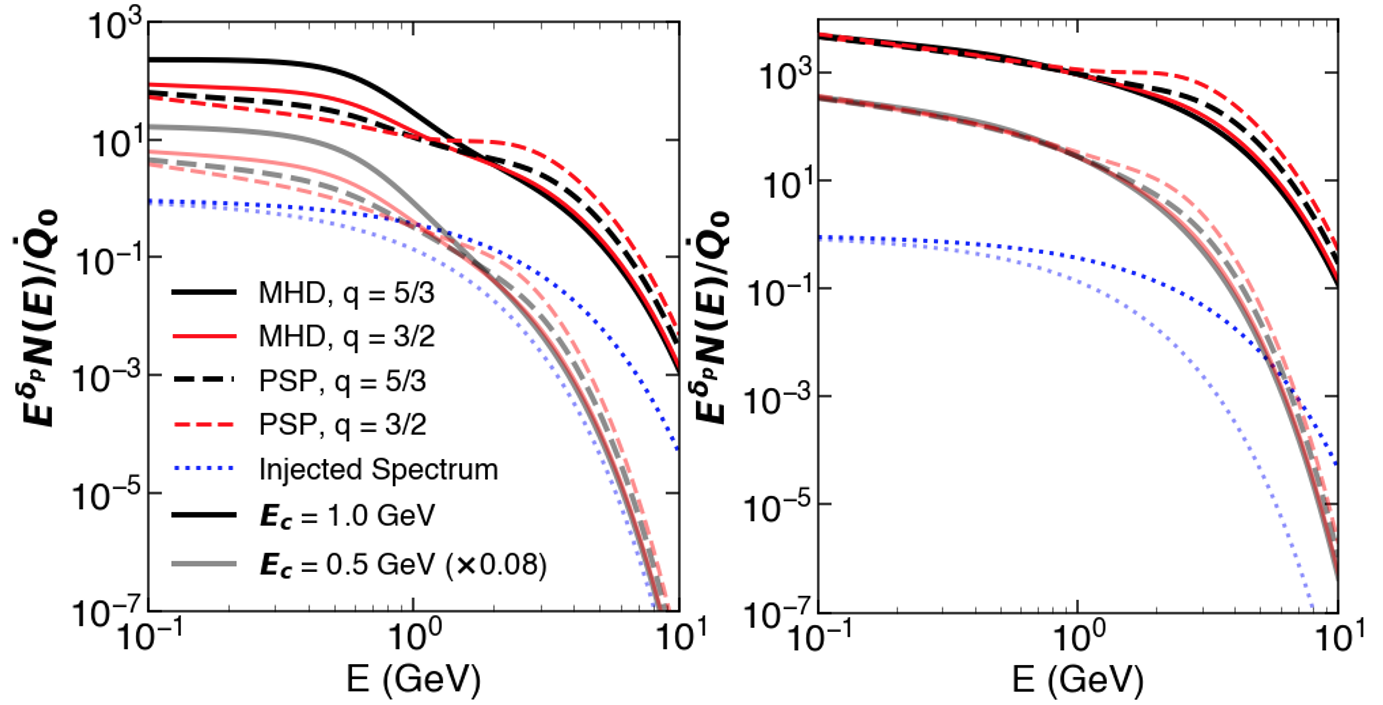}
		\caption{\mj{The variation of the proton spectra near the photosphere for a CME-shock at $\sim$2~\Rs (left) and $\sim$11~\Rs (right), for injected spectra power-law index $\d_p=4.0$, and for $q=3/2$ (red) and 5/3 (black) and power-law cutoff energy $E_c=0.5$~GeV (light curves) and $E_c=1.0$~GeV (dark curves).}}
		\label{fig:spec1}
	\end{figure}

\vahe{With this injected spectrum, Eq.~(\ref{injectedspec}) gives the proton spectrum at any point between the CME-shock and the Sun. Figure~\ref{fig:spec1} shows the spectrum at the Sun, where the gamma-rays are produced, for the shock at $r_{sh}\simeq 2$~\Rs\ (left) and $r_{sh}\simeq 11$~\Rs\ (right). The upper set of curves is for $E_c = 1$~GeV and the lower set for $E_c = 0.5$~GeV. Within each set the four transport models ($q=5/3$ and $3/2$, with $\tau_{p,p}$ from the MHD simulation and from the PSP extrapolation) are shown. The blue dotted curves show the injected spectra. Here $\d_p$ and $E_c$ are held fixed, although both may in reality vary with $r_{sh}$ as the shock weakens.}
\footnote{\vahe{
${\dot Q}_0$ is a dimensioned rate  related to the total rate of injected particles above some minimum energy $E_{\rm min}\sim 1$~MeV; ${\dot  Q}(E>E_{\rm min})={\dot Q_0}E_{\rm min}^{(1-\d)}(E_{\rm min}/E_c)\mathcal{E}_{\d_p}$, where the error function ${\cal E}_{\d_p}\sim 1/(\d_p-1)$ for $E_{\rm min}/E_c\sim 10^{-3}\ll 1$.}
}
\vahe{The relative normalization of these spectra depends on the variation of the unknown injection rate ${\dot Q}(r_{sh})$with shock distance, and their absolute normalization  is uncertain because of the uncertainty in the degree of focusing of the proton into smaller areas as they reach the photosphere and below, where the $B$ fields may converge by a large degree.}
\vahe{The energy spectra of the four transport models at low energies are identical for $r_{sh}\sim 11$ \Rs, with index $\sim -(\d_p+1)$, and vary slightly for $r_{sh}\sim 2$ \Rs, with PSP models yielding a slightly steeper index, $\sim -(\d_p+0.5)$, consistent with gamma-ray spectral index of $\sim -4$ for $\d_p\sim 3.5$,which is close to what is used in Figuere~\ref{fig:spec1}.}

\section{Summary and Discussion}
\label{sec:summary}

\vahe{We have investigated the transport of protons accelerated at a CME-driven shock from the downstream region of the shock back to the solar surface. The work is motivated by \f-LAT observations of long-duration gamma-ray emission from flares associated with fast CMEs, in which the emission centroid drifts across the solar disk and several events originate in active regions behind the limb (BTL) as seen from Earth. Using the global MHD simulation of the 2014 September 1 BTL event with the AWSoM model \citep{jin18}, we obtained the spatial and temporal evolution of the key plasma parameters -- magnetic field strength, Alfv\'en velocity, turbulence energy density, and magnetic field scale height -- along the field lines connecting the shock to the Sun. These parameters allowed us to compute the accelerated protons scattering mean free path, $\lambda_{mfp}$, as they scatter with MHD turbulence, and the ratio of the scattering time, $\tsc=\lambda_{mfp}/v$, to the characteristic time scale, $\tau_{p,p}$, given by Eq.~(\ref{tscfits}), shown in Figure \ref{fig:tauppandscattimes}. Because the simulated turbulence level is uncertain, we also obtained an independent estimate of $\tau_{p,p}$ by extrapolating power-law fits to PSP measurements of the magnetic field and turbulence energy density inward from $r\gtrsim 0.1$~au. In addition to the energy density of the turbulence, this calculation requires its spectral parameters, primarily the spectral index $q$ in the inertial range $k_{\rm min}<k<k_{\rm max}$, for which we used the Kolmogorov ($q=5/3$) and Iroshnikov--Kraichnan ($q=3/2$) values, and the injection scale $k_{\rm min}$, which we parameterized as $k_{\rm min}=1/(\xi H_B)$ with $\xi<1$ constrained by PSP observations. With $\tsc$ in hand, Eq.~(\ref{tescape}) gives the escape, or {\it residence}, time $\tesc(r,E)$: the time a proton of energy $E$, injected behind the shock, spends in the downstream region before reaching the Sun. The energy dependence of $\tsc$ and $\tesc$ near the Sun is shown in Figures~\ref{fig:scat1} and \ref{fig:tesc}, respectively.}

Our results show that, due to significant MHD turbulence downstream of the shock, protons can be efficiently scattered into the narrow loss cone and transported to the photosphere, overcoming objections based on freely streaming particles and strongly converging field geometries. \vahe{As evident from Fig.~\ref{fig:scat1}, $\tsc<\tau_{\rm cross}\sim H_B/v=(2/\d)(r/$\Rs)$\sim(r/$\Rs)~s, for $\d\sim 2$ and $v=c$, at high energies for MHD models, but almost all energies for PSP models, especially for $q=3/2$. This indicates we are mostly dealing with strong diffusion where field convergence has small effect on the transport of the protons. This is also evident from Fig.~\ref{fig:tesc}, showing that $\tesc\gg \tau_{\rm cross}$, especially when the shock is further away from the Sun.}

\citet{hutchinson2022} found that turbulence-driven scattering in the downstream region of the CME shock enhances particle precipitation. However, when the scattering mean free path becomes sufficiently small, outward advection by the solar wind becomes highly efficient, preventing protons from reaching the photosphere. \vahe{Whether precipitation or outward advection wins is determined by the ratio of the residence time $\tesc$ to the time over which the downstream flow carries the protons away from the Sun, $t_{\rm dyn}=\Delta r/u$, where $u$ is the flow speed behind the shock. Since $u\le v_{CME}$, $t_{\rm dyn} \ge r/v_{\rm CME}\sim 700(r/$\Rs)~s for CME velocity of 1000 km/s. For the shock at 2.1~\Rs, $t_{\rm dyn}\sim 1500$~s exceeds $\tesc$ at all energies for all models; for the shock at 11.3~\Rs, $t_{\rm dyn}\sim 7900$~s exceeds $\tesc$ except for the PSP model with $q=3/2$ in the range $2<E<30$~GeV. Diffusive transport to the photosphere is therefore faster than both outward advection and the evolution of the shock over most of the parameter space, so that precipitation is only minimally suppressed and the steady-state treatment neglecting the advection term in Eq.~(\ref{FPeq}) is justified. The advection-dominated regime identified by \citet{hutchinson2022} would apply when $\tesc$ becomes comparable to or longer than $t_{\rm dyn}$; a time-dependent treatment including the advection term is deferred to future work.}

We then explored the modification of the accelerated proton spectrum during transport. Assuming a power-law with exponential cut off injected spectrum  (consistent with both SEP and gamma-ray observations), we calculated how the proton spectrum evolves from the shock to the photosphere. Our results indicate that for the CME-shock and downstream conditions of the flare of 2014 September 1, a substantial fraction of accelerated protons can reach the photosphere with a spectrum similar to those required by \f observations.

\mj{Overall, our modeling of proton transport behind the CME-driven shock of the 2014 September 1 event -- based on a steady-state treatment of the transport equation and approximate analytic calculations of the relevant timescales, approximations whose validity we have verified against the derived dynamical and advection timescales and against the numerical transport solutions of \citet{effenberger18} -- supports the viability of the scenario in which the long-duration gamma-ray emission is produced by CME-driven shock accelerated protons reaching the surface of the Sun via transport from the downstream shock region.} While some aspects, such as the detailed comparison with SEP observations and the type II–gamma-ray correlation, require further investigation, our results provide a physically consistent framework to explain key observational features, including gamma-ray centroid drifts and emission from BTL flares, that are difficult to reconcile with alternative models such as acceleration in extremely large coronal loops. Note that the main aim of our study is to test the feasibility of transport back to the Sun and not to prove that this is the answer to all aspects of Fermi-LAT flare observations. The proof will require both the extension of characteristics of the CME environment to farther distances, and a more accurate treatment of the proton transport. Both of these are beyond the scope of the current paper. In addition, our results are terminated at a height of $\sim 10^9$ cm. Below this height the structure of the magnetic field becomes complicated including closed loop and complex open field lines. Transport in the lower corona and into the underlying chromosphere is currently receiving close scrutiny, largely motivated by efforts to address a related problem tied to another \f-LAT observation, namely gamma-ray emission from the quiet Sun (i.e., in the absence of flare activity). This emission is attributed to {\it galactic cosmic ray} protons (and other ions) interacting with solar material at and below the chromosphere, and its interpretation likewise requires modeling particle transport through a highly tangled magnetic field geometry. See, for example, the most recent treatments by \citet{dorner2026, li2024} and references therein.

Since there is no energy loss or gain during the transport (no change in ``entropy") the problem is reversible. The spectrum of protons at the photosphere, obtained from the observed gamma-rays, can be ``de-evolved" back to the CME using the same algorithm. Comparison of these spectra with the in situ observations of SEPs  can shed light on their transport from the CME to the instruments.

\mj{Finally, it is also useful to describe \textit{qualitatively} the particle time-intensity profile that a spacecraft encountering a CME-driven shock like the one studied here would measure — a measurement now within reach of Parker Solar Probe and Solar Orbiter. Before shock passage, the spacecraft would detect the protons escaping the upstream side of the shock (a typical SEP event). After shock passage the spacecraft enters the turbulent downstream region modeled here, and our results predict an elevated intensity, and slow diffusion of efficiently trapped particles and an energy-dependent evolution of the downstream spectrum due to the energy dependence of the escape or resident time, $\tesc$ (see Figures~\ref{fig:scat1} and \ref{fig:tesc}). In situ measurements of the downstream spectral evolution during near-Sun shock crossings would thus provide a direct test of the transport scenario proposed here.}

\begin{acknowledgments}
V.P. and M.J. conducted this study under the support of NASA grant 80NSSC21K1327. M.J. acknowledges support from NASA’s SDO/AIA contract (NNG04EA00C) to LMSAL. The simulation results were obtained using the Space Weather Modeling Framework (SWMF), developed at the Center for Space Environment Modeling (CSEM), University of Michigan (\url{https://github.com/MSTEM-QUDA/SWMF}). We are thankful for the use of the NASA Supercomputer Pleiades at Ames and its helpful staff for making it possible to perform the simulations presented in this paper.
\end{acknowledgments}

\end{document}